\documentclass[a4paper,11pt,twoside,reqno]{amsart}

\usepackage{bbm}
\usepackage{color}
\usepackage{graphicx}
\usepackage{amsmath,amsthm,amssymb,verbatim,amsfonts}
\usepackage{tensor}
\usepackage{slantsc}
\usepackage{slashed}

\usepackage{mathrsfs}
\usepackage{mathtools}
\usepackage{abstract}

\usepackage{bm}

\usepackage[sort]{cite}

\usepackage{hyperref}
\hypersetup{
    colorlinks=true,
    linkcolor=blue,
    filecolor=magenta,      
    urlcolor=cyan,
    citecolor=cyan,
}

\usepackage{enumerate}


\usepackage{etoolbox}
\makeatletter
\patchcmd{\@maketitle}{\newpage}{}{}{} 
\makeatother

\usepackage{mathtools}

\usepackage{marginnote}

\DeclareFontFamily{OT1}{rsfs}{}
\DeclareFontShape{OT1}{rsfs}{m}{n}{ <-7> rsfs5 <7-10> rsfs7 <10-> rsfs10}{}
\DeclareMathAlphabet{\mycal}{OT1}{rsfs}{m}{n}

{\catcode `\@=11 \global\let\AddToReset=\@addtoreset}
\AddToReset{equation}{section}

\newcounter{mnotecount}[section]

\newcommand{\R}{\mathbb R}

\newcommand{\supp}{\operatorname{supp}}

\theoremstyle{plain}
\newtheorem{thm}{Theorem}[section]
\newtheorem{lem}[thm]{Lemma}
\newtheorem{prop}[thm]{Proposition}

\theoremstyle{definition}

\newtheorem{rem}[thm]{Remark}

\numberwithin{equation}{section}

\newcommand{\vertiii}[1]{{\left\vert\kern-0.25ex\left\vert\kern-0.25ex\left\vert #1 
    \right\vert\kern-0.25ex\right\vert\kern-0.25ex\right\vert}}

\newcommand{\mnote}[1]
{\protect{\stepcounter{mnotecount}}$^{\mbox{\footnotesize
$
\bullet$\themnotecount}}$ \marginpar{
\raggedright\tiny
$\!\!\!\!\!\!\,\bullet$\themnotecount: #1} }

\newcommand{\tsig}{\widehat{\Sigma}}
\newcommand{\tsigp}{\widehat{\Sigma}_+}
\newcommand{\tsign}{\widehat{\Sigma}_-}
\newcommand{\nhat}{\widehat{N}}

\newcommand{\p}{\partial}

\newcommand{\absol}[1]{|#1|}
\newcommand{\absolg}[1]{|#1|_g}

\newcommand{\fR}{\mathfrak{R}}
\newcommand{\fRhat}{\widehat{\mathfrak{R}}}

\begin{document}


\title[Future attractors of Bianchi types II and V cosmologies with massless Vlasov matter]{Future attractors of Bianchi types II and V cosmologies with massless Vlasov matter}
\author[\textsc{H.~Barzegar}]{\textsc{Hamed Barzegar}}

\date{\today}


\maketitle
\begin{abstract}

It is shown that  the generalized Collins--Stewart radiation and Milne solutions are attractors of the massless Einstein--Vlasov system for Bianchi types II and V spacetimes, respectively.  The proof is based on an energy method and bootstrap argument which are used to determine the decay rates of the perturbations away from the attractors.
\end{abstract}

\section{Introduction}\label{Sec: intro}
Understanding the dynamics and the ultimate fate of cosmological models is one of the main goals of the mathematical cosmology which involves the study of the properties of cosmological solutions to the Einstein equations. A starting point is to consider those solutions with symmetries. However, the concordance model of cosmology is based on the solutions with the highest spatial symmetry assumptions, i.e., the isotropy and homogeneity, which is described locally by the FLRW metric. Although this model suggests very successful explanations and predictions for cosmological observations, it faces
several observational challenges (see e.g.~\cite{BuchertColeyKleinertRoukemaWiltshire} and references therein). A plausible strategy is therefore to drop at least the assumption of isotropy and examine the spatial homogeneous (SH) spacetimes that include two main classes: Bianchi models and Kantowski--Sachs spacetimes.
This program has been undertaken on a mathematical level for vacuum (e.g.~\cite{Ringstrom-Cauchy, RingstromBookTopology}) and for various matter models such as perfect fluid (see \cite{Coley-Book, EllisWainwright} for extensive introduction and results), scalar field (e.g.~\cite{Coley-Book, FajmanHeisselMaliborski}), other matter models (e.g.~\cite{RendallBook, NormannHervik2020} and references therein), and finally the collisionless kinetic gas which is the subject of the present work. On the observational level, however, it has been shown that \textsc{Planck} data do not favour the inclusion of a Bianchi type VII$_h$ component \cite{PlanckData}, although the observed large-scale intensity pattern mimics only that of
the Bianchi type  VII$_h$ because the Bianchi type  VII$_h$ is the most general SH spacetime that contains flat and open FLRW models for $h=0$ and $h \neq 0$, respectively.

A universe containing ensembles of self-gravitating collisionless particles is modeled by the Einstein--Vlasov system (EVS) which models stars, galaxies or even clusters of galaxies by collisionless particles and provides a realistic model for the universe at large scale (cf.~\cite{Andreasson2002, RendallIntro1996, RendallBook}) . In \cite{Rendall96} Rendall started the study of SH cosmologies with Vlasov matter and in \cite{RendallTod99} Rendall and Tod analyzed the SH cosmologies, with locally rotationally symmetry (LRS), by dynamical systems methods because of the reduction of the Einstein equations 	to a system of ODEs in SH spacetimes, although the transport (Vlasov) equation remains a PDE in general. The latter approach has been taken for various types of SH spacetimes (cf.~\cite{RendallUggla2000, HeinzleUggla06, CalogeroHeinzle2009, CalogeroHeinzle2010, CalogeroHeinzle2011, Heissel2012, FajmanHeissel2019}). Although the dynamical systems method provides global stability results it relies on the additional symmetries such as LRS and it cannot be applied to all types of SH cosmologies or to the cosmological models without spatial homogeneity. However, it is possible to relax extra symmetry assumptions on spacetime at the cost of assuming smallness of initial data to be able to prove the future stability of the EVS within the SH class. This has been accomplished for the massive EVS (cf.~\cite{NungesserI2010,NungesserIIV0,NungesserDiagonal2013,NungesserBianchiV}). The small-data assumption is needed due to the absence of the cosmological constant in the case of EVS whose existence was shown in \cite{Rendall93Cencorship}.  On the contrary,  in the presence of a positive cosmological constant the asymptotic behaviour of both massive and massless EVS is known  for all SH spacetimes except Bianchi type IX and  Kantowski--Sachs models, without assuming small data and additional symmetry assumptions \cite{Lee2004}.  The reason is that in the presence of the cosmological constant the perturbations have faster decay rates.

However, it turns out that the proof of future stability of the massless EVS without cosmological constant within SH spacetimes is even more challenging since the massless particles indicate slower decay in the course of the expansion of the universe.  Nevertheless, the stability (isotropization) of the massless EVS within Bianchi type I symmetry with small initial data has been proved recently \cite{BarzegarFajmanHeissel2020} (cf.~also \cite{LeeNungesserTod20}) and the purpose of the present work is to extend the previous results to the Bianchi types II and V models. We note that Bianchi type II  is the simplest case after the Bianchi type I in Bianchi class A and the Bianchi type V  is the simplest case in Bianchi class B.

In this paper, we show that the generalized Collins--Stewart (GCS) radiation solution is the future attractor of the massless EVS with Bianchi type II symmetry. This contradicts the expectation that the attractor of the massless EVS with the Bianchi type II symmetry should be the radiation perfect fluid solutions with the same symmetry (known as the Collins--Stewart (CS) solution; cf.~\cite{EllisWainwright}). 
The reason is the anisotropic behaviour of the Vlasov matter in this particular Bianchi type.
The GCS solution generalizes the CS solution to include anisotropic matter (cf.~Section~\ref{Sec: CS solutions}).
This was shown first in \cite{CalogeroHeinzle2011} using dynamical systems by assuming the LRS and reflection symmetry (see Section~\ref{Sec: reflection symmetry} for the precise definition). Here, we drop the assumption of LRS and prove not only the future stability for small perturbations of the  GCS solutions within the symmetry class, but also we show their asymptotic stability, i.e., we show that the  GCS solution is the future attractor. On the other hand, we show the future stability, in particular, the isotropization of small perturbations of the Milne solution within the Bianchi type V symmetric solutions to the massless EVS in full generality and show their asymptotic stability.  

The main challenge to proof the stability of the massless EVS in the Bianchi types II and V is to obtain  sufficiently fast decay estimates for the components of the energy-momentum tensor. To this end, we follow two different strategies for two different Bianchi types considered in this work. For the Bianchi type II we derive an evolution equation for the spatial energy-momentum tensor, based on which we estimate its decay. The reflection symmetry assumption is crucial for the desired estimate. For the non-diagonal Bianchi type V we rather exploit the properties of the Vlasov equation, in particular, the fact that the distribution function is  constant along the characteristic curves of the Vlasov equation. Then, similar to \cite{BarzegarFajmanHeissel2020} we use an energy method, which in conjunction with the bootstrap argument, not only provides the proof of future stability, but also determines the decay rates of the perturbations away from their fixed points. This method is more convenient and stronger compared to methods used in the previous works on massive EVS and is expected to be applicable to the class of inhomogeneous solutions.

Let us briefly discuss the geometry and its role.  The Bianchi type II geometry corresponds to Nil geometry in Thurston's classification. Therefore, it seems natural to expect that one can prove the stability of the massless EVS with the Bianchi type VI$_0$ symmetry which corresponds to the Sol geometry in Thurston's classification. Unfortunately, we were not able to show its stability by the method used here (cf.~Remark~\ref{rem: on estimate in vi0}).  Bianchi type V, on the other hand, corresponds to the hyperbolic geometry.  Hence, by Mostow's rigidity theorem one would already expect that these models isotropize since we consider compact spatial topology (cf.~\cite{BarrowKodama2001}).  However,  compactness plays no role in our analysis and we provide the decay rates at which the perturbations tend to zero and thus our work extends the results of \cite{NungesserBianchiV} to the massless case.
Moreover,  Bianchi type V cosmologies generalize the negatively curved FLRW models, in particular,  it contains the Milne model as a special case for which several robust nonlinear stability results have been established without any symmetry assumptions for vacuum \cite{AnderssonMoncrief2011},  for the EVS \cite{AnderssonFajman2020},  for the Einstein--Maxwell and other systems \cite{BrandingFajmanKroencke2019},  and more recently for the Einstein--Vlasov--Maxwell system in \cite{BarzegarFajman2020}.

The paper is organized as follows: Section \ref{Sec: Preliminaries} focuses on the preliminaries, in particular, notations, a brief introduction into the Bianchi models,  ($3+1$)-decomposition of the Einstein equations, the notion of reflection symmetry,  a short introduction into the Vlasov matter and EVS in Bianchi types II and V, and finally introducing the attractors. In Section \ref{Sec: main results} we formulate the main theorems of the paper. Finally, in Section \ref{Sec: proof} we first make precise our notion of smallness and bootstrap assumptions and then we provide estimates for the energy-momentum tensor and the Hubble parameter. Then, the energy functions are introduced which enables us to
 prove the main theorems of the paper.

\section{Preliminaries}\label{Sec: Preliminaries}

\subsection{Notation}\label{Sec: notation}

In this paper, we consider a Lorentzian manifold $\overline{M} = I \times M$, with an open interval $I \subset \R$ and $M$ being a three-dimensional Riemannian manifold that is also a simply connected three-dimensional Lie group. The Lorentzian and Riemannian metrics are denoted by $\bar{g}$ and $g$ with their associated covariant derivatives $\overline{\nabla}$ and $\nabla$, respectively.  Greek indices stand for the spacetime coordinates on $\overline{M}$ whereas Latin indices denote the spatial coordinates on $M$. 
We occasionally use the notation $\absolg{\cdot}$ for the norm of a vector field or a symmetric tensor with respect to a given metric $g$. 
Finally, $C$ denotes any positive constant which is uniform in the sense that it does not depend on the solution. However, the value of $C$ may change from line to line.

\subsection{Bianchi models} 

In this section, we briefly introduce the Bianchi models and refer the reader to \cite{EllisWainwright} for a detailed discussion. The Bianchi models are a subclass of SH spacetimes whose isometry group has three-dimensional subgroup $M$ which acts simply transitively on the spacelike orbits that foliate the Lorentzian manifold $\overline{M}$ (cf.~Section~\ref{Sec: notation}) and are called \emph{surfaces of homogeneity}. Therefore, in Bianchi models we have $\overline{M} = I \times M$. Let $M_t$ be the Lie group $M_t:= \{t\} \times M$ which admits a left-invariant frame $\{E_i\}$ with its dual $\{W^i\}$. 
Moreover, Bianchi models admit a Lie algebra of Killing vector fields $K_1$, $K_2$, and $K_3$ which are tangent to the orbits of the group and satisfy the commutation relation
$$
 [K_i, K_j] = \tensor{C}{_{ij}^k} \, K_k
 \,,
$$
where $\tensor{C}{_{ij}^k}$ are the \emph{structure constants}.
Let $E_0$ be a unit vector field  normal to the group orbits. In general, we have $[E_\mu, E_\nu] = \gamma^\alpha_{\mu \nu} \, E_\alpha$ where $\gamma^\alpha_{\mu \nu}$ are the \emph{commutation functions}.  Then, one has a natural choice for the time coordinate such that it commutes with the left-invariant frame $\{E_i\}$. This has the consequence that the commutation functions are constants and can be identified with the structure constants by a constant linear transformation, i.e.,
$$
 [E_i, E_j] =   \tensor{C}{_{ij}^k} \, E_k
 \,.
$$
In this way $\{E_i\}$ and hence $\{W^i\}$ are time-independent and the metric in the left-invariant frame reads
$$
 \bar{g} = - dt^2 + g_{i j} (t) W^i W^j
 \,.
$$
Bianchi models can be classified by classifying the structure constants. The structure constants can be decomposed uniquely as
$$
 \tensor{C}{_{ij}^k}
 =
 \epsilon_{i j \ell} n^{k \ell}
 +
 a_i \delta^k_j
 -
 a_j \delta^k_i
 \,,
$$
where $n^{ij}$ is a constant symmetric matrix and $a_i$ are constant. There are two Bianchi classes based on this decomposition; Bianchi class A where $a_i = 0$ and Bianchi class B where $a_i \neq 0$ for at least one $i \in \{1,2,3\}$. Moreover, in Bianchi class A trace of the structure constant vanishes, i.e., $\tensor{C}{_{ij}^j} = 0$. Finally, by a unimodular transformation we can diagonalize the matrix $n^{ij}$ as $n^{ij} = \text{diag}(\hat{n}_1, \hat{n}_2, \hat{n}_3)$ (cf.~e.g.~\cite{Ringstrom-Cauchy}). Therefore, one can classify Bianchi class A only by $\hat{n}_i$ whereas for Bianchi class B one needs also $a_i$. 

\begin{rem}
 The $1$-forms $W^i$ which are dual to the frame $\{E_i\}$ satisfy the Maurer–Cartan equation
$$
 dW^i = - \frac{1}{2}  \tensor{C}{_{jk}^i} W^j \wedge W^k
 \,,
$$
which allows us to determine the $1$-forms in terms of local coordinates for each Bianchi type (cf.~Table~8.2. in \cite{ExactSolutions}). In the Bianchi types studied in the present paper the $1$-forms have the form
\begin{align*}
  \{dx - z dy, dy, dz\} &\quad \text{for Bianchi type II}\,,
  \\
  \{dx, e^x dy, e^x dz\} &\quad \text{for Bianchi type V}
  \,. 
\end{align*}
\end{rem}
\begin{rem}
Because a cosmological spacetime is supposed to be the one which admits a compact Cauchy hypersurface,  we consider the \emph{locally} spatially homogeneous spacetimes which constitute a broader class of spacetimes that admit a compact Cauchy hypersurface. The reason is that if we restrict ourselves to the Bianchi models that admit a compact Cauchy hypersurface only Bianchi types I and IX should be considered. A spacetime is locally spatially homogeneous if each point has a neighbourhood that is isometric to an open subset of a spatially homogeneous spacetime. For more technical discussion we refer to \cite{Rendall95GlobalProperties}.
\end{rem}

\vspace{.5em}
\subsection{(3+1)-decomposition}
In the frame introduced in previous section, we define (cf.~\cite{RendallBook})
$$
 \rho := T^{00}
 \,,
 \quad
 j_i := T^0_i
 \,,
 \quad
 S_{ij} := T_{i j}
 \,,
$$
which are the energy density, matter current density, and the spatial part of the energy-momentum tensor, respectively, in terms of the energy-momentum tensor $T_{\alpha \beta}$ which satisfies the Einstein equation $G_{\alpha \beta} = 8 \pi T_{\alpha \beta}$ where $G_{\alpha \beta}$ is the Einstein tensor.
We write the Einstein equations in ($3+1$)-decomposition in the Hamiltonian and momentum constraint equations
$$
\begin{aligned}
 R -  k^{ij} k_{ij} + k^2 
 &= 
 16 \pi \rho
 \,,
 \\
 \nabla^j k_{i j}  
 &= 
 8  \pi j_i
 \,,
\end{aligned}
$$
and the evolution equations
\begin{subequations}
\begin{align}
 \p_t g_{i j} 
 &=
 - 2 k_{i j}
 \,,
 \\
 \p_t k_{i j} 
 &=
 R_{i j}
 +
 k \, k_{i j}
 -
 2 k_{i \ell} k^\ell_j
 -
 8 \pi
 \left(
  S_{i j} - \tfrac{1}{2} S g_{ij}
 \right)
 -
  4 \pi \rho g_{ij}
\,.\label{eq: evolution for k}
\end{align} 
\end{subequations}
Here, $R_{ij}$ is the Ricci tensor associated to the metric $g_{ij}$, $R=g^{ij} R_{ij}$ is the Ricci scalar, $k_{ij}$ and $k = g^{ij} k_{ij}$ are the second fundamental form of the hypersurface of spatial homogeneity $M_t$ and its trace, respectively, and finally $S=g^{i j} S_{ij}$. Another useful evolution equation is achieved by taking the trace of \eqref{eq: evolution for k} and using the Hamiltonian constraint
\begin{equation}\label{eq: k}
 \p_t k 
 =
 k_{i j} k^{ij}
 +
 4 \pi (S + \rho )
 \,.
\end{equation}

Further, we introduce variables that are usually used in SH cosmological models. First, we decompose the second fundamental form as 
\begin{equation}\label{eq: decomposition of kij}
 k_{ij} = \sigma_{ij} - H g_{i j}
\end{equation}
to introduce $\sigma_{ij}$ which is the trace-free part of $k_{ij}$ and is referred to as \emph{shear tensor} in cosmological context,  and the Hubble parameter by 
$$
H:= - \tfrac{1}{3} k
\,.
$$
\begin{rem}\label{rem: time origin}
Since the future geodesic completeness is known for these models \cite{Rendall95GlobalProperties}, one can choose the time origin without loss of generality as $t_0 = - 3 [\beta k(t_0)]^{-1}$, where $\beta > 0$ (cf.~\eqref{eq: beta} below) depends on the specific fixed point under study. This requires to assume that $k<0$ for all time. This is indeed the case for the Bianchi models in the present work \cite{Rendall93Cencorship}. Thus, $H>0$ for all time and $t_0 = [ \beta H(t_0) ]^{-1}$. 
\end{rem}
We next define the Hubble normalized variable
$$
 \Sigma^i_j := H^{-1} \sigma^i_j
 \,,
$$
with the evolution equation
\begin{equation}\label{eq: evolution for Sigma ij general}
 \p_t \Sigma^i_j
 =
 - H
  \left[
   \left( 3 - \p_t(H^{-1}) \right) \left( \Sigma^i_j - \delta^i_j \right)
   -
   \frac{1}{H^2} \tensor{R}{^i_j}
   +
   \frac{8 \pi}{H^2} \tensor{S}{^i_j}
   -
   \frac{4\pi}{H^2} (S - \rho) \delta^i_j
  \right]
  \,.
\end{equation}
It turns out that the computations will be simplified if one works with
$$
\begin{aligned}
 \Sigma_+
 &:=
 - \tfrac{1}{2} 
  \left(
   \Sigma^2_2 + \Sigma^3_3
  \right)
  \,,
 \\
 \Sigma_-
 &:=
 - \tfrac{1}{2 \sqrt{3}} 
  \left(
   \Sigma^2_2 - \Sigma^3_3
  \right)
 \,, 
\end{aligned}
$$
because $\Sigma^i_j$ is trace-free.  Note that $\Sigma^1_1 = 2 \Sigma_+$, $\Sigma^2_2 = - \Sigma_+ - \sqrt{3} \Sigma_-$, and $\Sigma^3_3 = - \Sigma_+ + \sqrt{3} \Sigma_-$. Moreover, in the diagonal case we have $\absolg{\Sigma}^2 = 6 (\Sigma_+^2 + \Sigma_-^2)$.
We also introduce the simplifying notations 
$$
 \Omega := \frac{8 \pi \rho}{3 H^2}
\,, 
$$
and
\begin{equation}\label{eq: q}
 q
 :=-1- \frac{\dot{H}}{H^2},
\end{equation}
which are called \emph{density parameter} and  \emph{deceleration parameter},  respectively,  in cosmological context.
Finally, we define 
$$
\tensor{\mathfrak{R}}{^i_j} : =  \frac{1}{ 6 H^2} \tensor{R}{^i_j}
\,,
\quad
\text{with}
\quad
\mathfrak{R} := \frac{1}{ 6 H^2} R
\,.
$$

\subsection{Massless Vlasov matter}

We give a quick introduction to the Vlasov matter in the context of the general relativity and we refer to \cite{Andreasson2002, RendallIntro1996, RendallBook} for more details.

In this paper, we assume that  all particles are massless and modeled by a distribution function which is defined on the mass shell
$$
 \mathcal{P}
 :=
 \{
  (x,p) : |p|_{\bar{g}}^2 = 0,  \,  p^0 <0
 \}
 \subset T\overline{M}
 \,,
$$
with $p=p^\mu \p_\mu$ and $p^\mu$ being the canonical coordinates on the tangent bundle of $\overline{M}$. Then,  one can associate an energy-momentum tensor to a distribution function $\bar{f}: \mathcal{P} \rightarrow [0, \infty)$ by
$$
 T^{\alpha \beta} [\bar{f}] 
 :=
 \int_{\mathcal{P}_x} \bar{f}(x,p) p^\alpha p^\beta \,  d\mu_{\mathcal{P}_x}
 \,,
$$
where $d\mu_{\mathcal{P}_x}$ is the Riemannian measure induced 	on $\mathcal{P}_x$ 	by the Lorentzian metric $\bar{g}$ at a given point $x$,  and is given by
$$
 d \mu_{\mathcal{P}_x}
 :=
 \frac{\sqrt{|\det \bar{g}|}}{- p_0} dp^1 \wedge dp^2 \wedge dp^3
 =
 \frac{1}{- p_0} d\mu_p
 \,, 
$$
where $d\mu_p := \sqrt{\det g} \, dp^1 \wedge dp^2 \wedge dp^3$. Note that if the metric $g$ is diagonal then $-p_0 = p^0 = \absolg{p}$.  

Instead of dealing with $\bar{f}$ we consider $f : = \bar{f} \circ \rm{pr}^{-1}$ which is referred to as \emph{distribution function} for the remainder of the paper. Here,  the projection map is defined by $\rm{pr} : \mathcal{P} \rightarrow T \overline{M}$ with $(x, p^0, p^i) \mapsto (x, p^i)$.

Then, the transport equation reads
$$
 p^\mu \p_\mu f
 -
\overline{\Gamma}^i_{\mu \nu} \, p^\mu p^\nu \p_{p^i} f
 =
 0
 \,,
$$
where the components of the connection are given by $\overline{\nabla}_{E_\alpha} E_\beta = \overline{\Gamma}^\lambda_{ \beta \alpha} E_\lambda$ and read
\begin{equation}\label{eq: connection components}
 \overline{\Gamma}^\lambda_{ \beta \alpha}
 =
 \frac{1}{2} \bar{g}^{\lambda \mu} 
 \left[
  E_\beta (\bar{g}_{\mu \alpha})
  +
   E_\alpha (\bar{g}_{\beta \mu})
  -
   E_\mu (\bar{g}_{\alpha \beta})
   +
   \gamma^\sigma_{\alpha \beta} \, \bar{g}_{\mu \sigma}
   +
   \gamma^\sigma_{\mu \alpha} \, \bar{g}_{\beta \sigma} 
   -
   \gamma^\sigma_{\beta \mu} \, \bar{g}_{\alpha \sigma}
 \right]
 \,.
\end{equation}
Moreover, we assume that $f(x, p^a)$ has compact support in momentum space at time $t_0$.

\subsection{Reflection symmetry}\label{Sec: reflection symmetry}

In the case of Bianchi type II symmetry we restrict ourselves to the solutions of the Einstein--Vlasov system  which admit the reflection symmetry, i.e., on the initial data we impose the following conditions (cf.~\cite{Rendall96})
$$
\begin{aligned}
 f_0(p_1, p_2, p_3)
 &= 
  f_0(p_1, - p_2, - p_3)
  =
   f_0(- p_1,- p_2, p_3)
 \,,
 \\
 g(t_0) 
 &=
 \text{diag}(g_{11}(t_0), g_{22}(t_0), g_{33}(t_0))  
 \,,
 \\
 k(t_0)
 &=
 \text{diag}(k_{11}(t_0), k_{22}(t_0), k_{33}(t_0))  
 \,.
\end{aligned}
$$
Under these conditions we have $j_i(t_0) = 0$ and $S_{i j}(t_0)$ is diagonal. The reflection symmetry is preserved by the evolution equations for all time, i.e., $g_{ij}$, $k_{ij}$, and $S_{ij}$ will remain diagonal and $j_i = 0$ for all time. In general, it is true that initial data with symmetry yield solutions of the Einstein–Vlasov equations with symmetry (cf.~\cite{RendallBook}).

\subsection{The generalized Kasner exponents}

Let $\kappa_i$ be the eigenvalues of $k_{ij}$ with respect to $g_{ij}$, i.e., they solve the equation
\begin{equation}\label{eq: eigenvalue of k}
 \det 
  \left(
   \tensor{k}{^i_j} - \kappa \delta^i_j
  \right)
  =
  0
  \,.
\end{equation}
Then, the generalized Kasner exponents are given by
$$
 \mathfrak{p}_i
 :=
 \frac{\kappa_i}{\sum_ \ell \kappa_\ell}
 =
 - \frac{\kappa_i}{3H} 
 \,.
$$
This is a useful concept from which one can extract information, e.g., about the spacetime anisotropy.
However, in contrast to the Kasner exponents, the generalized Kasner exponents are, in general, functions of time $t$. Moreover, they satisfy the first Kasner condition automatically, i.e., $\sum_i \mathfrak{p}_i = 1$, but not the second one necessarily, i.e., the condition $\sum_i (\mathfrak{p}_i)^2 = 1$ could be violated. In the case of the Kasner spacetime the generalized Kasner exponents do coincide with the Kasner exponents.

\subsection{The massless Einstein--Vlasov system with Bianchi symmetry}\label{Sec: Einstein--Maxwell system}

Assuming that the distribution function $f$, which determines the energy-momentum tensor of the Vlasov matter,  is compatible with the Bianchi symmetry,  one obtains the transport equation in the following form
\begin{equation}\label{eq: Vlasov}
 \p_t f(t,p^a)
 +
 \left(
  2 \tensor{k}{^i_j} p^j
  -
  \absolg{p}^{-1} \, \tensor{\omega}{^i_{j k}} \, p^j p^k
 \right)
 \p_{p^i} f(t, p^a)
 =
 0
 \,,
\end{equation}
where $\tensor{\omega}{^i_{jk}}$ are the Ricci rotation coefficients defined in terms of the structure constants $\tensor{C}{_{jk}^i}$ of the Lie algebra by 
\begin{equation}\label{eq: Ricci toation}
 \tensor{\omega}{^i_{jk}}
 :=
 \frac{1}{2} g^{im}
 \left(
  C_{mjk}
  +
  C_{mkj}
  -
  C_{jkm}
 \right)
 \quad
 \text{with}
 \quad
 C_{ijk}
 :=
 g_{km} \, \tensor{C}{_{ij}^m}
 \,.
\end{equation}
This follows directly from \eqref{eq: connection components}. In general,  we have $ \tensor{\omega}{^i_{ij}} =0$ and $\tensor{\omega}{^i_{ji}} =  - \tensor{C}{_{j i}^i}$.  Moreover, in Bianchi class A we have $ \tensor{\omega}{^i_{ji}} = 0$.

The relevant components of the energy momentum tensor entering the Einstein equations then are (cf.~\cite{RendallBook})
$$
\begin{aligned}
 \rho &= \int_{\mathbb R^3\setminus\{0\}} f(t,p^a) \absolg{p} \, d\mu_p
 \,, 
 \\
 j_i
&=
\int_{\mathbb R^3\setminus\{0\}} f(t,p^a) p_i \,  d\mu_p
\,,
\\
S_{ij}&= \int_{\mathbb R^3\setminus\{0\}} f(t,p^a)  \frac{p_i p_j }{\absolg{p}} \, d\mu_p
\,,
\end{aligned}
$$
where the integrals exclude the element $p^a=0$ to assure regularity of the integrand. We drop the domain for simplicity in the following.  Note that in the massless case $S = \rho$. Therefore, the constraint and evolution equations take the form
\begin{subequations}
\begin{align}
\Omega
 &=
 1 -  \tfrac{1}{6} \absolg{\Sigma}^2 + \fR
 \,,\label{eq: Hamiltonian constraint}
 \\
 \p_t \Sigma^i_j
 &=
 - H
  \left[
   \left( 2 - q \right) \left( \Sigma^i_j - \delta^i_j \right)
   -
  6 \tensor{\fR}{^i_j}
   +
  8 \pi H^{-2} \tensor{S}{^i_j}
  \right]
  \,,\label{eq: evolution for Sigma ij}
  \\
 \p_t H
 &=
 - H^2 (1 + q)
 \,,\label{eq: evolution for H}
\end{align}
\end{subequations}
where
$$
  q 
 =
 1 + \fR + \frac{1}{6} \absolg{\Sigma}^2
 \,.
$$
Equation \eqref{eq: evolution for Sigma ij} can be written in terms of $\Sigma_+$ and $\Sigma_-$ for diagonal components
\begin{align*}
 \p_t \Sigma_+
 &=
 - H
  \left[
   \Sigma_+ (2 - q)
   +
   \Omega
   -
   2 \fR
   +
   3\left( \tensor{\fR}{^2_2} + \tensor{\fR}{^3_3} \right)
  \right]
  +
  \frac{4 \pi}{H}
   \left(
    \tensor{S}{^2_2}
    +
    \tensor{S}{^3_3}
  \right)
  \,,
  \\
  \p_t \Sigma_-
  &=
  - H
 \left[
  \Sigma_-
  \left( 2 - q  \right)
  -
 \sqrt{3}
  \left( 
    \tensor{\mathfrak{R}}{^3_3} - \tensor{\mathfrak{R}}{^2_2} 
  \right)
  \right]
  +
  \frac{4 \pi }{\sqrt{3} H}
  \left(
   \tensor{S}{^2_2}
   -
    \tensor{S}{^3_3}
  \right)
  \,.
\end{align*}
Extra evolution equations shall be coupled to the above system for analyzing the future stability that depend on the specific Bianchi model, i.e.,  the evolution equations of the following variable  that are used in diagonal case  in  Bianchi type II (cf.~\cite{Ringstrom-Cauchy} and \cite{CalogeroHeinzle2011}) (no summation on repeated indices in the following equations is assumed)
$$
 N_i := \frac{n_i}{H}
 \,,
 \quad
 \text{with}
 \quad
 n_i
 := 
 \hat{n}_i  \frac{g_{ii}}{\sqrt{\det g}} 
 \,,
$$
based on which one can express the Ricci tensor as
$$
 \tensor{R}{^i_i}
 =
 \frac{1}{2}
 \left[
   n_i^2
  -
  ( n_j -  n_k)^2 
 \right]
 \,,
$$
where $(ijk)$ denotes a cyclic permutation of $(123)$. In the Bianchi type V we consider the evolution equation for $\fR$ instead. 

In the following we summarize the concrete form of the variables defined until now for Bianchi types II and V.

\subsubsection{Bianchi type II}
In this case we have
$$
 \hat{n}_1 = 1
 \,,
 \quad
 \hat{n}_2 = 0
 \,,
 \quad
 \hat{n}_3 = 0
 \,,
$$
and the only non-vanishing components of the structure constants are
\begin{equation}\label{eq: structure constants II}
 \tensor{C}{_{2 3}^1} 
 =
 1
 =
 - \tensor{C}{_{3 2}^1} 
 \,.
\end{equation}
The components of the Ricci tensor in diagonal case are
$$
 \tensor{R}{^1_1}
 =
 -\tensor{R}{^2_2}
 =
 -\tensor{R}{^3_3}
 =
 -R
 =
 \frac{1}{2} n_1^2
 \,.
$$
Note that $\fR = - N_1^2/12$.  Then,  the evolution equations for $(\Sigma_+, \Sigma_-, N_1)$ in Bianchi type II with reflection symmetry read
\begin{subequations}\label{eq: system for II}
\begin{align}
 \p_t \Sigma_{+}
 &=
 - H
 \big[
  1
  +
  \Sigma_+
  \left(
   2 - q
  \right)
  - 
  \tfrac{5}{12} N_1^2
  -
  \left(
    \Sigma_+^2 + \Sigma_-^2 
  \right)  
 \big]  
  +
  \frac{4 \pi }{ H}
  \left(
  \tensor{S}{^2_2}
   +
   \tensor{S}{^3_3}
  \right)
 \,,
 \\
  \p_t \Sigma_{-}
 &=
  - H
  \Sigma_-
  \left(
   2 - q
  \right)
  +
  \frac{4 \pi }{\sqrt{3} H}
  \left(
   \tensor{S}{^2_2}
   -
    \tensor{S}{^3_3}
  \right) 
 \,,
\\
  \p_t N_1
  &=
  - H N_1
      \left(
        4 \Sigma_+ - q
      \right) 
    \,.
\end{align}
\end{subequations}

\subsubsection{Bianchi type V}
The Bianchi type V is of class B; this means that $a_1 \neq 0$. On the other hand, $\hat{n}_i = 0$ for $i=1,2,3$. The non-vanishing components of the structure constants are then
$$
\tensor{C}{_{1 j}^i}
=
a_1 \delta^i_j
=
- \tensor{C}{_{j 1}^i}
\,,
\quad
i, j \in \{ 2,3  \}
\,.
$$
Without loss of generality we set $a_1 = 1$. Recall that in this case we do not assume the reflection symmetry.
Moreover, we have (cf.~ Equation~(1.95) in \cite{EllisWainwright})
$$
  \tensor{R}{^i_j}
 =
 \frac{1}{3} R \delta^i_j
 \quad 
 \text{with}
 \quad
 R
 =
 -6 a_1 a^1 
 =
 -6 a_1^2 g^{11}
 =
 -6 g^{11}
 \,,
$$
because the trace-free part of $R_{ij}$ vanishes.
Then, 
$$
\tensor{\fR}{^i_j}
 =
 \frac{1}{3} \fR \delta^i_j
 \quad
  \text{with}
 \quad  
 \fR = - \frac{1}{H^2} g^{11}
\,,
$$
and the evolution equations read (cf.~\eqref{eq: evolution for Sigma ij})
\begin{subequations}\label{eq: system for V}
\begin{align} 
 \p_t \Sigma^i_j
 &=
 - H
  (2 - q) \Sigma^i_j
  -
   \frac{8 \pi}{H} \tensor{S}{^i_j}  
\,;
\quad
i \neq j
\,,
\\  
 \p_t \Sigma_+
 &=
 - H
  \left[
   \Sigma_+ (2 - q)
  \right]
  +
  \frac{4 \pi}{H}
   \left(
    \tensor{S}{^2_2}
    +
    \tensor{S}{^3_3}
  \right)
  +
  H \Omega
  \,,
  \\
  \p_t \Sigma_-
  &=
  - H
  \Sigma_-
  \left( 2 - q  \right)
  +
  \frac{4 \pi }{\sqrt{3} H}
  \left(
   \tensor{S}{^2_2}
   -
    \tensor{S}{^3_3}
  \right)
  \,,
  \\
  \p_t \fR
  &=
  2  H \fR  
  \left(
   2 \Sigma_+
   +
   q
  \right)
  -
  \frac{2}{H}
   \left(
    \Sigma^1_2 \, g^{12}
    +
    \Sigma^1_3 \, g^{13}
   \right)
  \,.
\end{align}
\end{subequations}
Note that in this case we pulled out $\Omega$   in the evolution equation for $\Sigma_+$ and did not exploit the Hamiltonian constraint because the energy density estimate is good enough to control the matter terms.

\vspace{1em}
\begin{rem}
 To all systems of evolution equations mentioned above, one should couple the evolution equation for the Hubble parameter $H$. Moreover, they are constrained by the Hamiltonian constraint \eqref{eq: Hamiltonian constraint} and they are coupled to the Vlasov equation through the components of the energy-momentum tensor.
\end{rem}

\subsection{The generalized Collins--Stewart solution}\label{Sec: CS solutions}

In \cite{CalogeroHeinzle2011} Calogero and Heinzle showed that  the Bianchi type II models having the LRS with massless collisionless particles  have a future fixed point with the metric
\footnote{We note a misprint in Equation~(103a) of \cite{CalogeroHeinzle2011} where $\gamma_2$ should read $\gamma_2 = \frac{3 + \beta + w (1 - \beta)}{\beta (1 - w) + 4 (1 + w)}$ which can be easily checked by using the coordinates of the corresponding fixed point $C_\flat$ given there.}
$$
\overline{g}_{\mathrm{GCS}}
=
-dt^2
+
 \mathfrak{g}_{ij}(t) W^i \otimes W^j
 \,,
 \quad
 \text{with}
 \quad
  \mathfrak{g}(t)
 =
 \text{diag}
  \big(
   \tfrac{8}{9} t^\frac{2}{3}
   \,,
   t^\frac{4}{3}
   \,,
   t^\frac{4}{3}
  \big)
\,.
$$
The dynamical variables at this fixed point read
\begin{equation}\label{eq: CS variables}
 H = \frac{5}{9} t^{-1}
 \,,
 \quad
 \Sigma_+ = \frac{1}{5}
 \,,
 \quad
 \Sigma_- = 0
 \,,
 \quad
 N_1 = \frac{6 \sqrt{2}}{5}
 \,,
 \quad
 q= \frac{4}{5}
 \,,
\end{equation}
and the generalized Kasner exponents are 
$$
 \mathfrak{p}_{\rm{GCS}}
 =
 \left(
  \tfrac{1}{5}, \tfrac{2}{5}, \tfrac{2}{5}
 \right)
 \,.
$$

\subsection{The Milne solution}
The Milne solution is a negatively curved FLRW model with vanishing energy density and pressure whose scale factor is $t$. The Milne model can be considered as a part of Minkowski spacetime described in comoving coordinates adapted to the wordline of a particle. 
Its metric in the left-invariant frame is given by
$$
\overline{g}_{\mathrm{M}}
=
-dt^2
+
 \mathbf{g}_{ij}(t) W^i \otimes W^j
 \,,
 \quad
 \text{with}
 \quad
  \mathbf{g}_{ij}(t)
 =
 t^2 \delta_{i j}
\,.
$$
The dynamical variables at this fixed point are
\begin{equation}\label{eq: Milne variables}
 H =  t^{-1}
 \,,
 \quad
 \Sigma^i_j = 0
 \,,
 \quad
 \fR =  -1
 \,,
 \quad 
 q=0
 \,,
\end{equation}
and the generalized Kasner exponents, as in other FLRW models, are 
$$
 \mathfrak{p}_{\rm{M}}
 =
 \left(
  \tfrac{1}{3}, \tfrac{1}{3}, \tfrac{1}{3}
 \right)
 \,.
$$

\vspace{.5em}
\section{Main results}\label{Sec: main results}

In this section we summarize the main theorems of the paper. 
\begin{thm}\label{Theorem 1}
Consider $C^\infty$ initial data $(g(t_0), \Sigma(t_0), f(t_0))$ for the massless Einstein--Vlasov system with reflection symmetric Bianchi type II symmetry at $t_0=5 \left[ 9 H(t_0) \right]^{-1}$. Assume that $\left| \Sigma_+(t_0) - \tfrac{1}{5} \right|$, $\left| \Sigma_- (t_0) \right|$, and $\left| N_1(t_0) - \tfrac{6 \sqrt{2}}{5} \right|$ are sufficiently small. Then,
\begin{equation}
\begin{aligned}
 \Sigma_+ - \tfrac{1}{5}
 &=
 \mathcal{O} \big(t^{- \frac{4}{15}} \big) 
 \,,
 \\
 \Sigma_- 
 &=
 \mathcal{O} \big(t^{- \frac{4}{15}} \big) 
 \,,
 \\
 N_1 -  \tfrac{6 \sqrt{2}}{5}
 &=
 \mathcal{O} \big(t^{- \frac{4}{15}} \big) 
 \,,
\end{aligned}
\end{equation}
for large time $t$. Moreover, the generalized Kasner exponents are
\begin{equation}
 \mathfrak{p}
 =
 \mathfrak{p}_{\rm{GCS}}
 +
 \mathcal{O} \big(t^{- \frac{4}{15}} \big) 
 \,.
\end{equation}

\end{thm}
\vspace{.4em}
\begin{thm}\label{Theorem 2}
 Consider $C^\infty$ initial data $(g(t_0), \Sigma(t_0), f(t_0))$ for the massless Einstein--Vlasov system with Bianchi type V symmetry at $t_0= \left[  H(t_0) \right]^{-1}$. Assume that $\left| \Sigma^i_j(t_0) \right|$ for all $i,j=1,2,3$, and $\left| \fR(t_0) + 1 \right|$ are sufficiently small. Then, there exists a positive constant $\epsilon \ll 1$ such that  for $ i,j = 1,2,3$,
\begin{equation}
\begin{aligned}
 \Sigma^i_j
 &=
 \mathcal{O} \big(t^{- 2 + \epsilon} \big) 
 \,,
 \\
 \fR + 1 
 &=
 \mathcal{O} \big(t^{- 2 + \epsilon} \big)
 \,,
\end{aligned}
\end{equation}
hold for large time $t$. Moreover, the generalized Kasner exponents are
\begin{equation}
 \mathfrak{p}
 =
 \mathfrak{p}_{\rm{M}}
 +
 \mathcal{O} \big(t^{- 2 + \epsilon} \big)
 \,.
\end{equation}
\end{thm}

\section{Proof of the main theorems}\label{Sec: proof}

In this section we provide the proof of the main theorems of the foregoing section. In the following, we suspend the summation convention for repeated indices on a tensor field or in an expression with more than two repeated indices.

\subsection{Bootstrap argument}\label{Sec: bootstrap}

We make use of bootstrap argument to prove the main theorems. We refer the  reader  to \cite{RendallBook} for a discussion and to \cite{TaoBook} for a rigorous proof of the bootstrap principle. In the following we say that the bootstrap assumptions hold if there are smooth solutions to the systems \eqref{eq: system for II} and \eqref{eq: system for V} such that on the time interval $[t_0, T)$ for some $T>t_0$ the following conditions hold:

\begin{enumerate}

\item in Bianchi type II
\begin{equation*}
\begin{aligned}
 \left|
   \Sigma_{+}
   -
   \tfrac{1}{5}
 \right|
 & \leq 
  C \left(  1 + t \right)^{-\alpha}
 \,,
 \\
  \left|
   N_1
   -
   \tfrac{6 \sqrt{2}}{5}
 \right|
 & \leq 
  C \left(  1 + t \right)^{- \alpha}
 \,,
 \\
 \left|
   \Sigma_-
 \right|
 & \leq 
  C \left(  1 + t \right)^{-\alpha}
 \,,
\end{aligned}
\end{equation*}
for $1/5 < \alpha < 4/15$, and

\item in Bianchi type V
\begin{equation*}
\begin{aligned}
\phantom{xxxxxxxxxxxxxxxxxxxx}
 \left|
   \Sigma^i_j
 \right|
 & \leq 
  C \left(  1 + t \right)^{-\alpha}
 \,,
  \quad
 \text{for}
 \quad
 i,j = 1,2,3 
 \\
  \left|
   \fR + 1
 \right|
 & \leq 
  C \left(  1 + t \right)^{- \alpha}
 \,,
\end{aligned}
\end{equation*}
for $ 2 < \alpha < 7/2 $.

\end{enumerate}

We then wish to show that $T=\infty$. This is done by improving the bootstrap assumptions in the sense that the decay rates would be higher than the bootstrap assumptions after analyzing the system of evolution equations. Then, a continuation criterion finishes the proof.

Note that in what follows $\alpha$ has the values given in this section for different Bianchi types.

\begin{rem}
 In the same way one shows the global existence which is already done by Rendall in \cite{Rendall93Cencorship}. Here, we rather examine the stability by the bootstrap argument.
\end{rem}

\begin{rem}
 The bootstrap assumptions reflect the small-data assumptions. That is, the initial data for the massless EVS are $\varepsilon$-close to the GCS and Milne solutions in Bianchi types II and V cases, respectively.
\end{rem}

\vspace{.5em}
\subsection{Linearization}\label{Sec: linearization}

We first linearize the systems of evolution equations from Section~\ref{Sec: Einstein--Maxwell system} around their future attractors. For this purpose we introduce
$$
 \mathcal{O}(|\widehat{X}|)
 := 
     \mathcal{O}(\absolg{\tsig})
     +
    \mathcal{O}(|\nhat_1|) 
    +
    \mathcal{O}(|\fRhat|)
   \,,
$$
where the hatted variables are the shifted ones which are defined in the next sections. In Bianchi type V the second term does not exist. We use also $\mathcal{O}(|\widehat{X}|^2)$ to denote the  sum of all mixed second order terms. 
\subsubsection{Bianchi type II}
We linearize the system \eqref{eq: system for II} around the GCS solution by shifting the system to the origin using the variables (cf.~\eqref{eq: CS variables})
$$
 \widehat{\Sigma}_+ := \Sigma_+ - \frac{1}{5}
 \,,
 \quad  
 \tsign:=\Sigma_-
 \,,
 \quad   
 \nhat_1 := N_1 - \frac{6 \sqrt{2}}{5}
 \,.
$$
Hence,
$$
 q 
 =
 \frac{4}{5} + \frac{2}{5} \tsigp  - \frac{\sqrt{2}}{5} \nhat_1 + \tsigp^2 + \tsign^2 - \frac{1}{12} \nhat_1^2
 \,, 
$$
and
\begin{subequations}\label{eq: linearized II}
\begin{align}
 \p_t \tsigp
 &=
  - H
  \left(
    \tfrac{18}{25}  \tsigp
    -
    \tfrac{24 \sqrt{2}}{25} \nhat_1  
  \right)  
   +
  \frac{4 \pi }{ H}
  \left(
  \tensor{S}{^2_2}
   +
   \tensor{S}{^3_3}
  \right)
  +
  H \cdot \mathcal{O}(|\widehat{X}|^2)
 \,,
 \\
 \p_t \tsign
 &=
 - \frac{6}{5} H \tsign
 +
  \frac{4 \pi }{\sqrt{3} H}
  \left(
   \tensor{S}{^2_2}
   -
    \tensor{S}{^3_3}
  \right) 
  +
   H \cdot \mathcal{O}(|\widehat{X}|^2)
 \,,
\\
 \p_t \nhat_1
  &=
   - H
   \left(
    \tfrac{12}{25}  \nhat_1
    +
    \tfrac{108 \sqrt{2}}{25}  \tsigp
   \right) 
   +
   H \cdot \mathcal{O}(|\widehat{X}|^2)
    \,.
\end{align}
\end{subequations}

\subsubsection{Bianchi type V}

We linearize the system \eqref{eq: system for V} around the Milne solution by shifting the system to the origin using the variables (cf.~\eqref{eq: Milne variables})
$$
 \tsig^i_j : = \Sigma^i_j
 \,,
 \quad
 \fRhat := \fR + 1
 \,.
$$
Thus,
$$
 q
 =
 \fRhat + \frac{1}{6} \absolg{\tsig}^2
 \,,
$$
and
\begin{subequations}\label{eq: linearized system for V}
\begin{align} 
 \p_t \tsig^i_j
 &=
 - 2 H  \tsig^i_j
  -
   \frac{8 \pi}{H} \tensor{S}{^i_j}  
   +
   H  \mathcal{O}(|\widehat{X}|^2)
\,;
\quad
i \neq j
\,,
\\  
 \p_t \tsigp
 &=
 - 2 H
   \tsigp 
  +
  \frac{4 \pi}{H}
   \left(
    \tensor{S}{^2_2}
    +
    \tensor{S}{^3_3}
  \right)
  +
  H \Omega
   +
   H  \mathcal{O}(|\widehat{X}|^2)
  \,,
  \\
  \p_t \tsign
  &=
  - 2 H
  \tsign
  +
  \frac{4 \pi }{\sqrt{3} H}
  \left(
   \tensor{S}{^2_2}
   -
    \tensor{S}{^3_3}
  \right)
   +
   H  \mathcal{O}(|\widehat{X}|^2)
  \,,
  \\
  \p_t \fRhat
  &=
 - 2  H 
  \left(
   \fRhat
   +
   2 \tsigp
  \right)
  -
  \frac{2}{H}
   \left(
    \Sigma^1_2 \, g^{12}
    +
    \Sigma^1_3 \, g^{13}
   \right)
  +
   H  \mathcal{O}(|\widehat{X}|^2) 
  \,.
\end{align}
\end{subequations}

\vspace{.5em}
\subsection{Estimate of the Hubble parameter}

Before we give an estimate for the Hubble parameter we need to specify the constant $\beta$ and hence the time origin $t_0$ discussed in Remark~\ref{rem: time origin}.  We define
\begin{equation}\label{eq: beta}
 \beta
 :=
  \begin{cases}
  \frac{9}{5}
  \,;
  \quad
  \text{in Bianchi type II}
  \,,
  \\
  1
  \,;
  \quad
  \text{in Bianchi type V}
  \,.
 \end{cases}
\end{equation}
Then, the Hubble parameter has the following estimate.
\begin{lem}\label{lem: Estimate of H}
 Assume that the bootstrap assumptions hold. Then,
\begin{equation}\label{eq: estimate of H}
 H
 =
 (\beta t)^{-1}
 \left[
  1 + \mathcal{O}(t^{- \alpha})
 \right]
 \,,
\end{equation}
with $\beta$ defined in \eqref{eq: beta}.
\begin{proof}
The evolution equation for the Hubble parameter \eqref{eq: evolution for H} can be rewritten as
$$
\p_t (H^{-1})
 =
 1+q
 =
 \beta + \mathcal{O}(|\widehat{X}|)
 \,,
$$
where $q$ is written in terms of the shifted variables given in the previous section.
Hence,
$$
\beta
 -
 C
   t^{-\alpha}
 \leq
 \partial_t \left( H^{-1}  \right)
 \leq
 \beta
 +
  C
   t^{-\alpha}
  \,.
$$
Integrating this inequality on $[t_0, t)$ and using the time origin mentioned in Remark~\ref{rem: time origin} with $\beta$ defined in \eqref{eq: beta} completes the proof.
\end{proof}
\end{lem}

\vspace{.5em}
\subsection{Estimate for the metric components}

In this subsection we summarize the estimate of the metric components in different cases. We use the evolution equation for $g^{ij}$, i.e.,
\begin{equation}\label{eq: evolution for gii}
 \p_t g^{i j}
 =
 2H
 \left(
  \Sigma^i_\ell
  -
  \delta^i_\ell
 \right) g^{\ell j}
 \,.
\end{equation}
\begin{lem}\label{lem: estimate of gij}
 Assume that bootstrap assumptions hold. Then, for Bianchi type II we have
\begin{align}
 g^{11}
 &=
 t^{- \frac{2}{3}} 
 \left[
  \mathcal{G}^{11} + \mathcal{O}(t^{-\alpha})
 \right]
 \,,
 \\
 g^{ii}
 &=
 t^{-  \frac{4}{3}} 
 \left[
  \mathcal{G}^{ii} + \mathcal{O}(t^{-\alpha})
 \right]
 \,;
 \quad
 i=2,3
 \,,
\end{align}
whereas in Bianchi type V we have
\begin{equation}\label{eq: gij Bianchi V}
 g^{ij} 
 = 
 t^{-2}
 \left[
  \mathcal{G}^{i j} + \mathcal{O}(t^{-\alpha})
 \right]   
 \,.
\end{equation}
Here, $\mathcal{G}^{i j}$ are some positive constants.
\begin{proof}
Let us start with Bianchi types II. From \eqref{eq: evolution for gii} and Lemma~\ref{lem: Estimate of H} we get
\begin{align*}
 \p_t g^{11}
 &=
  2 H (2 \Sigma_+ - 1 ) g^{11}  
  \\
  &=
  2 H  \left(2 \tsigp + \tfrac{2}{5} - 1 \right) g^{11}   
  \\
  &\leq
  - \frac{6}{5} \beta^{-1}  t^{-1} g^{11} 
  +
  C t^{-1-\alpha} g^{11}
  \,,
\end{align*}
where we imposed the bootstrap assumptions. Recall that in Bianchi type II $\beta = 9/5$. Hence,
\begin{equation}
 \p_t \left( t^{\frac{2}{3}} g^{11} \right)
 \leq
 C t^{-1-\alpha} \left( t^{\frac{2}{3}} g^{11} \right)
 \,.
\end{equation}
Integrating this inequality on $[t_0, t)$ yields the claim of the lemma.

Similarly, for $g^{ii}$ with $i=2,3$ we find
\begin{align*}
 \p_t g^{ii}
 &=
 2 H 
  \left(
   -\Sigma_+ \pm \sqrt{3} \Sigma_- - 1
  \right)
  g^{ii}
  \\
  &=
  - 2 H 
  \left(
     \tfrac{6}{5} + \tsigp \pm \sqrt{3} \tsign
  \right)
  g^{ii}
  \\
  &\leq
 - \frac{4}{3} t^{-1} g^{ii} + C t^{-1-\alpha} g^{ii}
  \,. 
\end{align*}
Integrating this inequality on $[t_0, t)$ gives the estimate for $g^{22}$ and $g^{33}$.

For Bianchi type V we first look at the diagonal components
\begin{align*}
 \p_t g^{ii}
 &=
 2 H 
  \left(
  \Sigma^i_i - 1
  \right)
  g^{ii}
  \\
  &=
  2 H 
  \left(
  \tsig^i_i - 1
  \right)
  g^{ii}
  \\
  &\leq
  - 2 \beta^{-1} t^{-1} g^{ii} + C t^{-1-\alpha} g^{ii}
  \,,
\end{align*}
for $i = 1, 2, 3$. Recall that $\beta = 1$ in Bianchi type V. Integrating the inequality on $[t_0, t)$ gives $g^{ii} = t^{-2} \left[ \mathcal{G}^{ii} + \mathcal{O}(t^{-\alpha}) \right] $. Then, for the off-diagonal components ($i \neq j$) we find
\begin{align*}
 \p_t g^{i j}
 &=
 2 H 
  \left(
    \Sigma^i_i  -1
  \right)
  g^{ij}
  +
  2 H \sum_{i \neq \ell} \Sigma^i_\ell \, g^{\ell j}
  \\
  &\leq
  - 2 H g^{i j} + 2 H \left| \tsig^i_\ell g^{\ell j} \right|
  \\
  & \leq
  - 2  t^{-1} g^{i j} + C t^{-1 - \alpha} g^{i j} + C t^{-3-\alpha}
  \,,
\end{align*}
where we used $|g^{\ell j}| \leq \left( g^{\ell \ell} g^{j j} \right)^{1/2}$ since $g^{ij}$ is a Riemannian metric. Finally, integrating the last inequality on $[t_0, t)$ and using the Gr\"onwall's inequality completes the proof of the lemma.
\end{proof}
\end{lem}

\vspace{.5em}    
\subsection{Estimate of components of the energy-momentum tensor}

We need to estimate the components of the energy-momentum that appear in the systems of the evolution equations. This can be done in several ways. However, it is crucial to achieve the best decay rate such that one can show the stability. Therefore, we begin this subsection by analyzing the behaviour of the momenta of the distribution function $f$ along the characteristic curve $P^i(t)$ of the transport equation \eqref{eq: Vlasov} which reads
$$
 \frac{d P^i}{dt}
 =
  2 \tensor{k}{^i_j} p^j
  -
  \absolg{P}^{-1} \, \tensor{\omega}{^i_{j k}} \, P^j P^k
  \,,
$$
where $P^i (\bar{t}) = \bar{p}^i$ for a given $\bar{t}$. Moreover, the characteristic curve $P_i(t)$ takes a simpler form
\begin{equation}\label{eq: characteristic pi}
 \frac{d P_i}{dt}
 =
  -
  \absolg{P}^{-1} \, \tensor{\omega}{^\ell_{m n}} \, P_j P_k g^{j m} g^{k n} g_{ i \ell}
  \,,
\end{equation}
where $P_i (\bar{t}) = \bar{p}_i$ for a given $\bar{t}$.  In the following $P^i$ and $P_i$ indicate that $p^i$ and $p_i$ are parameterized by $t$, respectively. We prove a result about the characteristic curve $P_i$ which will be used later to estimate the energy density.
\begin{lem}\label{lem: estimate of characteristic}
 Assume that bootstrap assumptions hold. Then, for the Bianchi types II and V and for large time $t$ we have
\begin{equation}
 \left| P_i \right|
 =
 \mathcal{P}_i + \mathcal{O}(t^{-\alpha})
 \,;
 \quad
 \text{for}
 \quad
 i=1,2,3
 \,,
\end{equation}
where $\mathcal{P}_i$ are positive constants.
\begin{proof}
Define the positive function $F_i := g^{ii} P^2_i$ for $i=1,2,3$ in Bianchi type II. Then, by \eqref{eq: evolution for gii} we find
\begin{equation}\label{eq: evolution for F}
 \p_t F_i 
 =
 2H (\Sigma^i_i - 1) F_i
 \,,
\end{equation}
where we used the characteristic equation and the structure constants \eqref{eq: structure constants II}.
%
%
Thus, using Lemma~\ref{lem: Estimate of H}, applying the bootstrap assumptions, integrating \eqref{eq: evolution for F} on $[t_0, t)$ and finally comparing the result with that of Lemma~\ref{lem: estimate of gij} yield the claim of the proof for the Bianchi type II. In a similar vein, we define $P^2 := g^{i j} P_i P_j$ for Bianchi type V. Then,
$$
 \frac{d}{dt} P^2 = \p_t g^{ij} P_i P_j
 =
 - 2 H P^2 + 2 H \Sigma^{i j} P_i P_j
 \,,
$$
where we again used the characteristic equation \eqref{eq: characteristic pi}. Repeating the same argument as before completes the proof.
\end{proof}
\end{lem}

Now, we are ready to estimate the components of the energy-momentum tensor.

\begin{prop}\label{prpo: estimate of S}
 Assume that bootstrap assumptions hold. Then,
\begin{equation}
 H \tensor{S}{^i_j}
 =
 t^{  - \lambda}
  \left[
   \tensor{\mathcal{S}}{^i_j}
   +
   \mathcal{O}(t^{-\alpha})
  \right]
  \,,
\end{equation}
where 
\begin{equation}
 \lambda
 =
 \begin{cases}
  \frac{4}{3}
  \,;
  \quad
  \text{in Bianchi type II for} \,\, i = j \in \{2,3 \}
  \,,
  \\
  3
  \,;
  \quad
  \text{in Bianchi type V for} \,\, i, j \in \{1,2,3 \}
  \,.
 \end{cases}
\end{equation}
Here, $\tensor{\mathcal{S}}{^i_j}$ are positive constants.
\begin{proof}
We start with the time derivative of $S^{i j}$. Note that $p^i$ are time-independent. Then,
$$
\p_t {S}^{i j}
 =
 \int
 \p_t f(t, p^a) \, \frac{p^i p^j}{|p|_g} \, d\mu_p
 +
 \int
 f(t, p^a) p^i p^j 	\partial_t \left( |p|_g^{-1}  \right) \, d\mu_p
 +
  \int
 f(t, p^a) \frac{p^i p^j}{|p|_g} \, \partial_t  (d\mu_p)
 \,.
$$
Using the Vlasov equation \eqref{eq: Vlasov} for the first term, integrating it by parts, and finally using
\begin{align*}
 \p_{p^i}
    \left(
      |p|^{-1}_g
    \right)
 &=
 -  \absolg{p}^{-3} \, p_i
 \,,
 &
 \partial_t \left( |p|_g^{-1}  \right)
 &=
 k_{ij} p^i p^j  \absolg{p}^{-3}
 \,,
 &
  \tensor{\omega}{^i_{jk}} p^j p^k p_i
   &=
   0
  \,,
  \\
  \p_{p^i}
   \left(
     \absolg{p}^{-2}
   \right)
  &=  
   -
   2  \absolg{p}^{-4} \, p_i
   \,,
   &  
    \partial_t  (d\mu_p)
  &=
  - k \,  d\mu_p
  \,,
\end{align*}
where we used $ \partial_t  \sqrt{g} = - k  \, \sqrt{g}$ in the last equality,  we arrive at
$$
  \p_t \tensor{S}{^i_j}
  =
 \p_t (g_{j \ell} \, S^{i \ell})
 =
 - 4 H \tensor{S}{^i_j}
 +
 2 H \Sigma^i_\ell \, \tensor{S}{^\ell_j}
 +
 \tensor{Y}{^i_j}
 -
 H 
 \tensor{Z}{^{ i}_{ j}}
 \,,
$$
where
\begin{align*}
 Y^{i j}
 & :=
 2 \tensor{C}{_{a b}^c}
 \int f(t, p^k)  \, g^{a (i} p^{j)} p^b p_c \absolg{p}^{-2}  \,  d\mu_p
 \,,
 \\
  \tensor{Z}{^{ i}_{ j}}
 &:=
 \Sigma_{m \ell}   \int f(t, p^k) \,   
  p^\ell p^m p^i p_j \absolg{p}^{-3}  \,  d\mu_p
 \,. 
\end{align*}
It is readily seen that in reflection symmetric case the following holds
$$
 \tensor{Y}{^i_i} = 0
 \,;
 \quad
 i = 1,2,3
 \,.
$$
On the other hand, we have
\begin{align*}
 \Sigma_{m \ell} \, p^m p^\ell
 &=
 \Sigma^1_1 \, p^1 p_1
 +
 \Sigma^2_2 \, p^2 p_2
 +
 \Sigma^3_3 \, p^3 p_3  
 \\
 &=
 2 \Sigma_+ p^1 p_1
 -
 \Sigma_+ (p^2 p_2 + p^3 p_3)
 -
 \sqrt{3} \Sigma_- (p^2 p_2 - p^3 p_3)
 \\
 &=
 2 \Sigma_+
 \left[
  \absolg{p}^2 
  - 
  \tfrac{3}{2}
   \left(
    p^2 p_2 + p^3 p_3
   \right) 
 \right]
 -
 \sqrt{3} \Sigma_- 
  \left(
   \absolg{p}^2 - p^1 p_1 - 2 p^3 p_3
  \right)
  \,.
\end{align*}
Then, for $i = 2,3$ we have
\begin{equation}\label{eq: derivative of S}
\begin{aligned}
 \p_t \tensor{S}{^i_i}
 &=
 - 4 H \tensor{S}{^i_i}
 -
 2 H \left( \Sigma_+ \pm \sqrt{3} \Sigma_- \right) \tensor{S}{^i_i}
 -
 H 
  \left(
    2 \Sigma_+ - \sqrt{3} \Sigma_-
  \right)  
 \tensor{S}{^i_i}
 \\
 & \quad
 +
 3 H \Sigma_+
  \int f(t,p^a) \frac{p^2 p_2 + p^3 p_3}{\absolg{p}^3} p^i p_i \, d\mu_p
  -
  \sqrt{3} H \Sigma_-
  \int f(t,p^a) \frac{p^1 p_1 + 2 p^3 p_3}{\absolg{p}^3} p^i p_i \, d\mu_p
 \,.
\end{aligned}
\end{equation}
Note that $\tensor{S}{^i_i}$ is non-negative.  Using  $p^2 p_2 + p^3 p_3 \leq \absolg{p}^2$, we estimate the fourth term
$$
 3 H \Sigma_+
  \int f(t,p^a) \frac{p^2 p_2 + p^3 p_3}{\absolg{p}^3} p^i p_i \, d\mu_p
  \leq
  3 |\tsigp| \tensor{S}{^i_i} 
  + 
  \frac{3}{5} \int f(t,p^a) \frac{p^2 p_2 + p^3 p_3}{\absolg{p}^3} p^i p_i \, d\mu_p
 \,, 
$$
The fifth term of \eqref{eq: derivative of S} can be estimated similarly. Hence,
\begin{align*}
 \p_t \tensor{S}{^i_i}
 &\leq
 - H
 \left[
    \tfrac{21}{5} + \tsigp
   +  
  \sqrt{3} (1 \pm 2) \tsign
 \right]
  \tensor{S}{^i_i}
  +
  \left( 3 |\tsigp| + \sqrt{3} |\tsign| \right) H  \tensor{S}{^i_i}
 \\
 &\leq
  - \tfrac{7}{3}  t^{-1} \tensor{S}{^i_i}
  +
  C t^{-1 - \alpha} \tensor{S}{^i_i}
 \,,
\end{align*}
where we used \eqref{eq: estimate of H} and imposed the bootstrap assumptions. 
Consequently, 
$$
 \p_t 
 \left(
  t^{\frac{7}{3}} \tensor{S}{^i_i}
 \right)
 \leq
 C t^{-1 - \alpha} 
  \left(
  t^{\frac{7}{3}} \tensor{S}{^i_i}
 \right)
 \,.
$$
Integrating this inequality on $[t_0, t)$ yields the claim of the proposition.

For the Bianchi type V we need another strategy because of the non-diagonality. 
Since $f(t_0, p^a)$ has compact support and $P_i(t)$ is uniformly bounded for large time $t$ (cf.~Lemma~\ref{lem: estimate of characteristic}), there exists a constant $C$ such that 
$$
 f(t, p_i) = 0
 \,;
 \quad
 \text{if}
 \quad
 |p_i| \geq C
\,, 
$$
for large time $t$. Now, choose an (spatial) orthonormal frame $\{ \tilde{e}_i \}$. We denote the quantities in the orthonormal frame by a tilde, i.e., $\tilde{p}:=(\tilde{p}_1, \tilde{p}_2, \tilde{p}_3)$ and $d\mu_{\tilde{p}}:= d\tilde{p}_1 d\tilde{p}_2 d\tilde{p}_3$. Then, in the orthonormal frame for Bianchi type V we have (cf.~\cite{Lee2004})
$$
 f(t, \tilde{p}) = 0
 \,;
 \quad
 \text{if}
 \quad
 |\tilde{p}_i| \geq C t^{- 1}
 \,,
$$
where we used \eqref{eq: gij Bianchi V}.
Hence, we find
$$
 \rho
 =
 \int_{ 0< |\tilde{p}_i| \leq C t^{- 1} }
   f(t, \tilde{p})  |\tilde{p}|_\delta \, d\mu_{\tilde{p}}
 \,.
$$
On the other hand, $f(t, \tilde{p})$ remains constant along the characteristics, i.e.,
$$
 f(t, \tilde{p}) 
 \leq
 \lVert f_0 \rVert
 :=
 \sup \{ f(t_0, \tilde{p}) : \tilde{p} \in \supp f(t_0, \cdot) \} 
 \,.
$$
We thus have
$$
 \rho
 \leq 
 C \lVert f_0 \rVert t^{-1} \int_{ 0< |\tilde{p}_i| \leq C t^{- 1} }
   d\mu_{\tilde{p}}
  \leq
  C t^{- 4}
  \,.
$$
Finally, by Lemma~\ref{lem: Estimate of H} and the fact that $\absol{\tensor{S}{^i_j}} \leq \rho$ we finish the proof.
\end{proof}
\end{prop}
\begin{rem}\label{rem: on estimate in vi0}

If we drop the reflection symmetry assumption on the distribution function the terms $\tensor{Y}{^i_j}$ do \emph{not} vanish for Bianchi type II.  Also the components of $\tensor{Z}{^i_j}$ need more careful analysis in that case. This causes the main difficulty in non-diagonal case.  If one could manage to obtain good estimates for these quantities one is able to generalize the result of the present work to the non-diagonal case. However, in Bianchi type VI$_0$, even in diagonal case, the  method used above for Bianchi type II does not provide a good decay rate (the decay rate would be $t^{-1}$).  In fact, the characteristics method gives a better decay rate (the decay rate would be $t^{- 7/4}$), but it is not sufficient to conclude the stability of Bianchi type VI$_0$.
\end{rem}
\begin{rem}
One could also take advantage of the continuity equation $\overline{\nabla}_\mu T^{\mu 0} = 0$.  This equation, however, does \emph{not} lead to a good decay rate for the energy density both in non-diagonal Bianchi types II and VI$_0$. Nevertheless, one gets the same result for the Bianchi type V. This is due to the geometrical property of the Bianchi type V as discussed in Section~\ref{Sec: intro}. Note also that this result coincides with the known fact about radiation-dominated FLRW universes with the energy density $\rho \sim a^{-4}(t)$ where $a(t)$ is the scale factor.
\end{rem}

\vspace{.5em}
\subsection{Estimate of the generalized Kasner exponents}

Before we prove the main part of the Theorems~\ref{Theorem 1} and \ref{Theorem 2}, we first prove the second part of the them.
Let $\widehat{\sigma}^i_j$ denote the shifted values of $\sigma^i_j$. Then,  from \eqref{eq: decomposition of kij} and \eqref{eq: eigenvalue of k} it follows
$$
 \det 
  \left[
   \widehat{\sigma}^i_j - \left( \kappa + H - B H \right) \delta^i_j
  \right]
 = 
 0
 \,,
$$
where $B$ is the shifted value which depends on the type of Bianchi symmetry and the components of $\sigma^i_j$. For instance, in Bianchi type II and for $\sigma^1_1$ we have $B_1 = \tfrac{2}{5}$. Therefore, $\widehat{\kappa}_i := \kappa_i + (1 - B_i) H$ are the eigenvalues of $\widehat{\sigma}^i_j$ and $\sum_i ( \widehat{\kappa}_i )^2 = \absolg{\widehat{\sigma}}^2 = H^2 \absolg{\tsig}^2$ holds. We finally find 
$$
 \mathfrak{p}_i
 =
- \frac{\kappa_i}{3H}
=
  \frac{1}{3}
   \left(
    1 - B_i
   \right)
  +
 \mathcal{O}(\absolg{\tsig})
 \,.
$$
This proves the second parts of the theorems.

\vspace{.5em}
\subsection{Energy method}

In this final subsection we define energy functions for the linearized systems of evolution equations.  The energy method was first applied on the Bianchi type I with massless Vlasov matter in \cite{BarzegarFajmanHeissel2020}. It has two significant advantages. First it makes the calculations simple and second it may be relevant in generalizing the method to the inhomogeneous case. 

Note that the energy functions defined below are non-negative and only vanish at the fixed points.

\subsubsection{Bianchi type II}

For the linearized system \eqref{eq: linearized II} we define the energy by
$$
 \mathbf{E}_{\rm{II}}
 :=
 c_1 \tsigp^2
 +
 \tsign^2
 +
 c_2 \nhat_1^2
 \,,
$$
where $c_i$ are some positive constants for $i=1,2$. Then,
\begin{align*}
 H^{-1} \p_t E_{\rm{II}}
 &=
 - \frac{36}{25} c_1 \tsigp^2
 -
 \frac{12}{5} \tsign^2
 -
 \frac{24}{25} c_2 \nhat_1^2
 +
 2 \sqrt{2}
  \left(
   \tfrac{24}{25} c_1 - \tfrac{108}{25} c_2
  \right)
  \tsigp \nhat_1
  \\
  & \quad
  +
  \frac{4 \pi }{H^2}
  \left(
   \tensor{S}{^2_2}
    +
    \tensor{S}{^3_3}
  \right)
   \tsigp 
   +
  \frac{4 \pi }{\sqrt{3} H^2}
  \left(
   \tensor{S}{^2_2}
   -
    \tensor{S}{^3_3}
  \right) 
   \tsign
  +
   \mathcal{O}(|\widehat{X}|^3)  
  \,.
\end{align*}
Now, we introduce a decay inducing positive constant $\mu$ such that
$$
 H^{-1} \p_t \mathbf{E}_{\rm{II}}
 =
 -\mu \mathbf{E}_{\rm{II}}
 + 
 Q_{\rm{II}}
 +
  \frac{4 \pi }{H^2}
  \left(
   \tensor{S}{^2_2}
    +
    \tensor{S}{^3_3}
  \right)
   \tsigp 
   +
  \frac{4 \pi }{\sqrt{3} H^2}
  \left(
   \tensor{S}{^2_2}
   -
    \tensor{S}{^3_3}
  \right) 
   \tsign
  +
   \mathcal{O}(|\widehat{X}|^3) 
  \,,
$$
with the quadratic form
$$
 Q_{\rm{II}}
 :=
  (\mu - \tfrac{36}{25}) c_1 \tsigp^2
 +
 (\mu - \tfrac{12}{5} ) \tsign^2
 +
 ( \mu - \tfrac{24}{25} ) c_2 \nhat_1^2
 +
 2 \sqrt{2}
  \left(
   \tfrac{24}{25} c_1 - \tfrac{108}{25} c_2
  \right)
  \tsigp \nhat_1
$$
We wish to choose the constants $\mu$, $c_1$, and $c_2$ such that the quadratic form $ Q_{\rm{II}}$ is negative semidefinite and $\mu$ attains its maximum possible value. One then finds that this can be done by choosing
$$
 \mu = \frac{24}{25}
 \,, \quad
 c_2 = \frac{2}{9} c_1
 \,,
 \quad
 \text{for}
 \quad
 c_1 > 0
 \,.
$$
Thus, imposing the bootstrap assumptions, using Lemma~\ref{lem: Estimate of H} and the results of  Proposition~\ref{prpo: estimate of S} we arrive at
\begin{align*}
  \p_t \mathbf{E}_{\rm{II}}
  &\leq
  - \frac{24}{25} H \mathbf{E}_{\rm{II}}
  +
  \frac{4 \pi }{H}
  \left(
   \tensor{S}{^2_2}
    +
    \tensor{S}{^3_3}
  \right)
   \tsigp 
   +
  \frac{4 \pi }{\sqrt{3} H}
  \left(
   \tensor{S}{^2_2}
   -
    \tensor{S}{^3_3}
  \right) 
   \tsign
  +
  H \mathcal{O}(|\widehat{X}|^3)  
  \\
  &\leq
  - \frac{8}{15} t^{-1} \mathbf{E}_{\rm{II}}
  +
  C t^{- 4/3 - \alpha} 
  +
  C t^{-1-3 \alpha}
  \,.
\end{align*}
Now, recall that $1/5 < \alpha < 4/15$. We integrate the following inequality on $[t_0, t)$
$$
 \p_t 
  \left(
   t^{8/15} \mathbf{E}_{\rm{II}}
  \right)
  \leq
   C t^{- 4/5 - \alpha} 
  +
  C t^{-7/15 -3 \alpha}
$$
and find
$$
 t^{8/15} \mathbf{E}_{\rm{II}}
 \leq
 t_0^{8/15} \mathbf{E}_{\rm{II}}(t_0)
 +
 C t_0^{1/5 - \alpha}
 +
  C t_0^{8/15 - 3 \alpha}
  \,,
$$
which implies
$$
 \mathbf{E}_{\rm{II}} \leq C t^{- 8 / 15}
 \,.
$$
This in turn means
$$
 \left| \tsigp \right| = \mathcal{O}(t^{- 4 / 15})
 \,,
 \quad
 \left| \tsign \right| = \mathcal{O}(t^{- 4 / 15})
 \,,
 \quad
 \left| \nhat_1 \right| = \mathcal{O}(t^{- 4 / 15})
 \,,
$$
which improves the bootstrap assumptions and closes the bootstrap argument, and therefore finishes the proof of Theorem~\ref{Theorem 1}.

\vspace{0.5em}
\subsubsection{Bianchi type V}

For the linearized system \eqref{eq: linearized system for V} we define
$$
 \mathbf{E}_{\rm{V}}
 :=
  \sum_{i \neq j} \left( \tsig^i_j \right)^2
  +
 c_1 \tsigp^2 
 +
 \tsign^2
 +
 c_2
 \fRhat^2
 \,,
$$
for some positive constants $c_1$ and $c_2$. Then, similar to the Bianchi type II case we find
\begin{align*}
 H^{-1} \p_t \mathbf{E}_{\rm{V}}
 =
 - \mu \mathbf{E}_{\rm{V}}
 +
 Q_{\rm{V}}
 +
 \frac{C'}{H^2} \tensor{S}{^i_j} \Sigma^i_j 
 -
 \frac{4}{H^2} \fRhat \left( \Sigma^1_2 g^{12} + \Sigma^1_3 g^{13} \right)
 +
 \mathcal{O}(|\widehat{X}|^3)
 \,,
\end{align*}
where $\mu$ is a positive constant, $C'$ is some constant and
$$
  Q_{\rm{V}}
 :=
 (\mu - 4) \sum_{i \neq j} \left( \tsig^i_j \right)^2
 +
  (\mu - 4) c_1 \tsigp^2
 +
  (\mu - 4) \tsign^2
 +
 (\mu - 4) c_2 \fRhat
 -
 8 \tsigp \fRhat
 \,.
$$
The quadratic form $ Q_{\rm{V}}$ is negative semidefinite by choosing
$$
 \mu \leq 4 - \frac{4}{\sqrt{c_2}}
 \,,
 \quad
 c_2 > 0
 \,,
 \quad
 c_1 c_2 > 1
 \,.
$$
One can choose $c_2$ sufficiently large such that $\mu = 4 - 2\epsilon$ with $0 < \epsilon \ll 1$. Hence, by the smallness assumptions and Lemmas~\ref{lem: Estimate of H} and \ref{lem: estimate of gij} we obtain
$$
 \p_t \mathbf{E}_{\rm{V}}
 \leq
 - \mu t^{-1} \mathbf{E}_{\rm{V}}
 +
 C t^{-1 - 2 \alpha}
 +
 C t^{- 3 - \alpha}
 \,.
$$
Integrating
$$
 \p_t 
 \left(
  t^\mu \mathbf{E}_{\rm{V}} 
 \right)
 \leq
 C t^{\mu - 1 - 2 \alpha}
 +
 C t^{\mu - 3 - \alpha}
$$
on $[t_0, t)$ and keeping in mind that $ 2 < \alpha < 7/2$, one finds
$$
 t^\mu \mathbf{E}_{\rm{V}} 
 \leq 
 t_0^\mu \mathbf{E}_{\rm{V}}(t_0)
 +
 C t_0^{\mu - 2 \alpha}
 +
 C t_0^{\mu - 2 - \alpha}
 \,,
$$
which implies 
$$
 \mathbf{E}_{\rm{V}}  \leq C t^{-\mu}
 \,,
$$
and consequently
$$
 \left| \tsig^i_j \right|
 =
 \mathcal{O}(t^{- 2 +\epsilon})
 \,,
 \quad
  \left| \fRhat \right|
 =
 \mathcal{O}(t^{- 2 +\epsilon})
 \,.
$$
This improves the bootstrap assumptions, closes the bootstrap argument, and hence completes the proof of the Theorem~\ref{Theorem 2}.

\vspace{1em}
\subsection*{Acknowledgements} The author would like to thank David Fajman for many useful comments and discussions.  This work was supported in part by the Austrian Science Fund (FWF) via the project \emph{Geometric transport equations and the non-vacuum Einstein flow} (P 29900-N27).

\bibliographystyle{plain}
\bibliography{Bianchi-II-Vl-massless-Vlasov.bib}

\vspace{0.5cm}
\textsc{Hamed Barzegar\\
Gravitational Physics,\\
Faculty of Physics, University of Vienna,\\
Boltzmanngasse 5, A-1090 Vienna}\\ 
\vspace{-0.3cm}\\
\texttt{Hamed.Barzegar@univie.ac.at}\\

\end{document}